\documentclass{article}
\usepackage{spconf,graphicx}
\usepackage{booktabs}
\usepackage{balance}

\usepackage{etoolbox}

\usepackage[table]{xcolor} 
\usepackage{subcaption} 
\usepackage{amsmath,amssymb} 
\usepackage{lipsum} 
\usepackage[sort]{cite} 
\usepackage{hyperref} 
\hypersetup{
    colorlinks=false,
    pdfborder={0 0 0},
    pdfborderstyle={/S/U/W 0},
}
\usepackage[tight-spacing=true]{siunitx} 
\usepackage{acronym}
\acrodef{DNN}{deep neural network}
\acrodef{SI-SDR}{scale-invariant source-to-distortion ratio}
\acrodef{SDR}{source-to-distortion ratio}
\acrodef{SIR}{signal-to-interference ratio}
\acrodef{TV}{time-varying}
\acrodef{TI}{time-invariant}
\acrodef{TSE}{target speaker extraction}
\acrodef{SS}{speech separation}
\acrodef{DSB}{delay-and-sum beamformer}
\acrodef{SDB}{superdirective beamformer}
\acrodef{MVDR}{minimum variance distortionless response}
\acrodef{MPDR}{minimum power distortionless response}
\acrodef{STFT}{short-time Fourier transform}
\acrodef{TCN}{temporal convolution network}
\acrodef{SCM}{spatial covariance matrix}
\acrodefplural{SCM}[SCMs]{spatial covariance matrices}
\acrodef{GLN}{global layer normalization}
\acrodef{DOA}{direction of arrival}
\acrodef{TF}{time-frequency}
\acrodef{SC}{single-channel}
\acrodef{MC}{multi-channel}
\acrodef{TD-SpeakerBeam}{time-domain SpeakerBeam}
\acrodef{TasNet}{time-domain audio separation network}
\acrodef{BF}{beamformer}
\acrodef{STFT}{short-time Fourier transform}
\acrodef{TSNF}{Temporal Spatial Neural Filter}
\acrodef{IRM}{ideal ratio mask}
\acrodef{AS}{angular separation}
\acrodef{BGTSE}[BG-TSE]{Beamformer-guided TSE}
\acrodef{AF}{angular feature}
\acrodef{IPD}{inter-phase difference}
\acrodef{ASR}{automatic speech recognition}
\acrodef{PIT}{permutation invariant training}

\title{Beamformer-Guided Target Speaker Extraction}
\name{Mohamed Elminshawi, Srikanth Raj Chetupalli, Emanu\"el A.~P.~Habets}
\address{International Audio Laboratories Erlangen$^\dag$\thanks{$^\dag$A joint institution of the Friedrich-Alexander-Universit\"{a}t Erlangen-N\"{u}rnberg (FAU) and Fraunhofer IIS, Germany.}, Am Wolfsmantel 33, 91058 Erlangen, Germany}

\begin{document}
%
\maketitle
\begin{abstract}

We propose a Beamformer-guided Target Speaker Extraction \acused{TSE} (BG-TSE) method to extract a target speaker's voice from a \acl{MC} recording informed by the \acl{DOA} of the target. The proposed method employs a front-end \acl{BF} steered towards the target speaker to provide an auxiliary signal to a \acl{SC} \ac{TSE} system. By allowing for time-varying embeddings in the \acl{SC} \ac{TSE} block, the proposed method fully exploits the correspondence between the front-end \acl{BF} output and the target speech in the microphone signal. Experimental evaluation on simulated multi-channel $2$-speaker mixtures, in both anechoic and reverberant conditions, demonstrates the advantage of the proposed method compared to recent \acl{SC} and \acl{MC} baselines. 


\end{abstract}

\acresetall

\vspace{-0.3em}
\section{Introduction}
\vspace{-0.5em}
Extracting a target speaker's voice from a multi-talker mixture is essential for many speech processing technologies, including conferencing, speaker verification, and \ac{ASR}. Thanks to the powerful modeling capabilities of \aclp{DNN}, recent \ac{SC} \ac{SS} \cite{Luo2019} and \ac{TSE} \cite{vzmolikova2019speakerbeam,Delcroix2020} methods have reached remarkable performance in anechoic scenarios. However, in reverberant environments, the performance of \ac{SC} \ac{SS}/\ac{TSE} methods degrades due to the smearing of the spectro-temporal characteristics caused by reverberation \cite{chen2017cracking}.

One approach to cope with the limitations of \ac{SC} methods is to exploit the spatial properties of the sources by employing multiple microphones. Conventional spatial filtering, using, e.g., the \ac{MVDR} \ac{BF}, has been extensively studied in the literature \cite{souden2009optimal}. However, to compute the \ac{BF} weights, it is often required to estimate the \acp{SCM} of the interfering signal (and the target signal for some variants), which is challenging in a multi-talker scenario. Neural-based \ac{MC} \ac{TSE} methods have also been proposed \cite{vzmolikova2019speakerbeam, Delcroix2020, gu2019neural, gu2020temporal, li2019direction}. For example, in \cite{vzmolikova2019speakerbeam}, an enrolment-based \ac{SC} \ac{TSE} method was utilized to estimate \ac{TF} masks, which are then used for computing the \acp{SCM} of a back-end \ac{BF}. A subsequent work \cite{Delcroix2020} investigated incorporating spatial features, e.g., \acp{IPD}, into an enrolment-based \ac{SC} \ac{TSE} system. In \cite{li2019direction}, a set of fixed \acp{BF} steered towards different directions were employed, followed by selection based on the correlation with an enrolment signal \cite{li2019direction}. However, the aforementioned \ac{MC} \ac{TSE} methods require an enrolment utterance from the target speaker, which might not be available in some applications. An alternative enrolment-free \ac{MC} approach was proposed in \cite{gu2019neural, gu2020temporal}, where the authors assumed knowledge about the target speaker's \ac{DOA} and extracted hand-crafted directional features, e.g., \acp{AF}, which are used as auxiliary information in a \ac{SC} \ac{TSE} framework. However, such hand-crafted features are not guaranteed to effectively model the directional information of the target speaker.


In this work, we propose an alternative approach to exploit the \ac{DOA} of the target speaker in a \ac{TSE} framework. In particular, we employ a front-end \ac{BF}, e.g., \ac{DSB}, steered towards the target speaker to provide an auxiliary signal to a \ac{SC} \ac{TSE} system. The front-end \ac{BF} output is synchronous with the direct-path signal of the target speaker and has a better \ac{SIR} compared to the microphone signal. This initial enhancement provides a clue about the target speaker which the \ac{SC} \ac{TSE} can leverage to identify and extract the target speaker. We refer to this approach as \ac{BGTSE}. This work is motivated by several \ac{SC} \ac{TSE} studies \cite{wang2020speaker, elminshawi2021informed, chen22e_interspeech} that demonstrated the advantage of having a correspondence between the auxiliary signal and the target speech in the mixture. Furthermore, we investigate extending the proposed method with a back-end \ac{BF}, similar to \cite{ochiai2020beam, chen22e_interspeech}. The proposed method is evaluated on simulated multi-channel 2-speaker mixtures and achieved better extraction performance than recent \ac{SC} and \ac{MC} baselines\footnote{Audio examples are available Online \url{https://www.audiolabs-erlangen.de/resources/2023-ICASSP-BGTSE}}.

\vspace{-0.3em}
\section{Target Speaker Extraction}
\label{sec:TSE}
\vspace{-0.6em}
\subsection{Signal Model}
Let ${\bf{y}} = [y^{(1)}, \dots, y^{(C)}]$ denote a $C$-channel microphone signal, where $y^{(c)} \in \mathbb{R}^{T}$ represents the time-domain waveform of the $c$-th microphone having a length of $T$ samples. We assume that each microphone captures speech from a target speaker $x_S^{(c)} \in \mathbb{R}^{T}$ and other interfering speech signal(s), represented by $x_N^{(c)} \in \mathbb{R}^{T}$, i.e., $y^{(c)} = x_S^{(c)} + x_N^{(c)}$. In this work, the goal is to estimate the reverberant target signal $x_S^{(c)}$ given the multi-microphone recording, where the target is specified by the \ac{DOA}. We note that the DOA is readily available in applications such as video conferencing (from the camera feed). 

\vspace{-0.6em}
\subsection{Overview of Enrolment-based SC TSE Method}
\label{sec:sc_TSE}

\begin{figure}[t!]
    \centering
    \includegraphics[height=3.7cm]{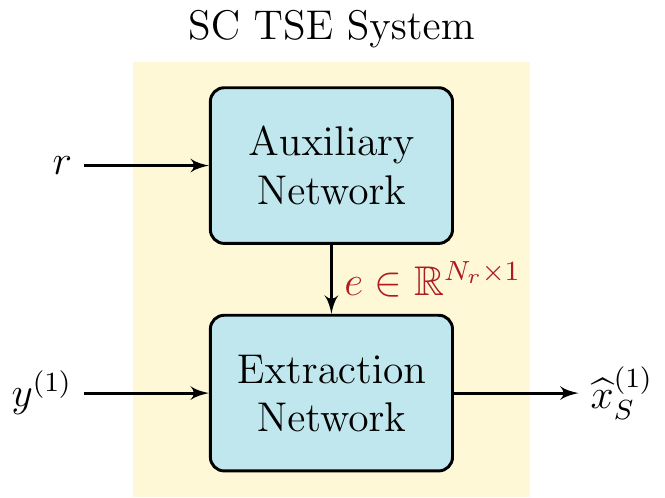}
    \caption{Enrolment-based SC \ac{TSE} system \cite{vzmolikova2019speakerbeam, Delcroix2020}. The auxiliary network maps the enrolment signal $r$ to a time-invariant (TI) embedding $e \in \mathbb{R}^{N_r \times 1}$ characterizing the target speaker.}
    \label{fig:sc_TSE}
    \vspace{-0.4em}
\end{figure}

The goal of \ac{SC} \ac{TSE} is to recover speech of the target speaker from an observed \ac{SC} mixture signal $y^{(1)}$ with the help of an auxiliary signal about that target, denoted by $r \in \mathbb{R}^{T_r}$, i.e., \mbox{$\widehat{x}^{(1)}_{S} = \text{TSE}(y^{(1)}, r)$}, where $T_r$ represents the length of the auxiliary signal in samples. Recent enrolment-based \ac{SC} \ac{TSE} methods \cite{vzmolikova2019speakerbeam, Delcroix2020} utilize two jointly trained networks: an auxiliary network and an extraction network, as shown in Figure~\ref{fig:sc_TSE}. The auxiliary network extracts target-specific features from the enrolment utterance and then aggregates them using a temporal pooling layer, resulting in an utterance-wise, i.e., \ac{TI} speaker embedding of dimension $N_r$, represented by $e \in \mathbb{R}^{N_r \times 1}$. The speaker embedding is then passed to the extraction network to inform it about the target speaker, i.e., $\widehat{x}^{(1)}_{S}=\text{Ext}(y^{(1)}, e)$.

\vspace{-0.6em}
\subsection{Proposed \ac{BGTSE} Method}
\label{sec:proposed}

\begin{figure}[t!]
    \centering
    \includegraphics[height=3.7cm]{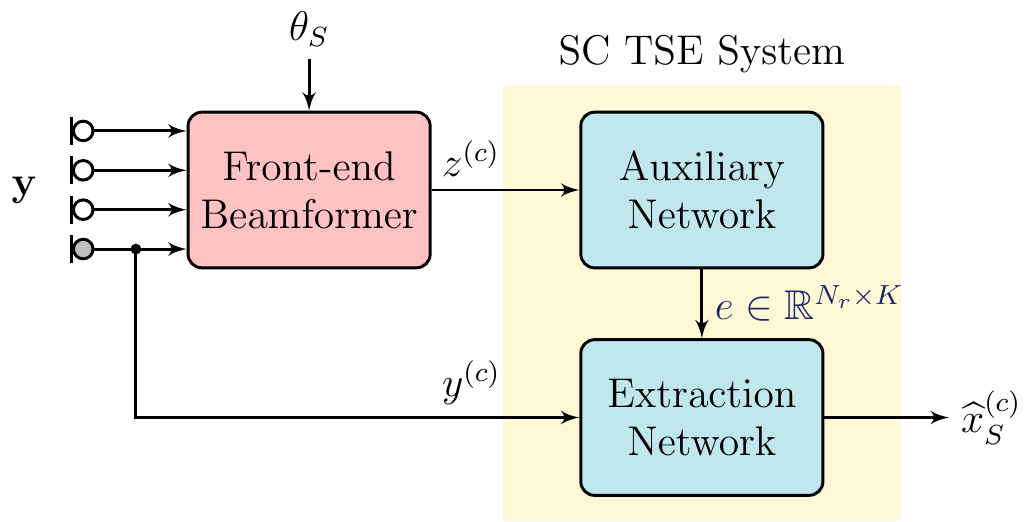}
    \caption{Proposed \acf{BGTSE} method. A front-end \ac{BF} steered towards the target speaker's direction $\theta_S$ provides an auxiliary signal to a \ac{SC} \ac{TSE} system employing time-varying (TV) embeddings $e \in \mathbb{R}^{N_r \times K}$.}
    \label{fig:proposed}
    \vspace{-0.4em}
\end{figure}

Here, we present our \acf{BGTSE} method, a simple \ac{MC} extension to the \ac{SC} \ac{TSE} system presented in Section~\ref{sec:sc_TSE}. As illustrated in Figure~\ref{fig:proposed}, a front-end \ac{BF} steered towards the target speaker's direction $\theta_S$ is employed, i.e.,  $z^{(c)} = \text{BF}_{\text{FE}}({\bf{y}},\,\theta_{S},\,c)$, where $z^{(c)} \in \mathbb{R}^T$ denotes the output of the front-end \ac{BF} and $c$ specifies the reference channel. The front-end \ac{BF} output is then used as an auxiliary signal for a \ac{SC} \ac{TSE} system, i.e., 
\vspace{-0.4em}
\begin{align}
    \label{eq:proposed}
	\widehat{x}^{(c)}_{S} = \text{TSE}(y^{(c)},\,z^{(c)}).
\end{align}
The role of the front-end \ac{BF} is to boost the signal coming from the desired direction, thereby accentuating the identity of the target speaker. Another advantage of using a front-end \ac{BF} is that it detaches the \ac{MC} input from the \acl{DNN}, thus allowing for better generalizability to different array geometries. In contrast to the system in Section~\ref{sec:sc_TSE}, we remove the last temporal pooling layer in the auxiliary network to allow for \ac{TV} embeddings, i.e., $e \in \mathbb{R}^{N_r \times K}$, where $K$ represents the number of time-frames. This way, the \ac{SC} \ac{TSE} system can better exploit the correspondence between the auxiliary signal, i.e., front-end \ac{BF} output, and the target speech in the mixture. Further details about the system configuration are provided in Section~\ref{sec:model_configs}. 
 
\vspace{-0.6em}
\subsection{Related Work}
The proposed \ac{BGTSE} method was inspired by several \ac{SC} \ac{TSE} \cite{wang2020speaker, elminshawi2021informed} works that demonstrated the advantage of having a correspondence between the auxiliary signal and the target speaker in the mixture, unlike enrolment utterances, as well as allowing for \ac{TV} embeddings. In \cite{wang2020speaker}, an initial estimate of the target speaker from a first-stage separation network was used as an auxiliary signal for a second stage. In \cite{elminshawi2021informed}, a \ac{SC} \ac{TSE} system was applied in an acoustic echo cancellation scenario, where the task was to extract the target component (echo signal) in a reverberant input mixture using its anechoic version (reference far-end signal) as an auxiliary signal. 

\vspace{-0.6em}
\subsection{Back-end Beamforming}
\label{sec:be_beamforming}

Neural-based \ac{TSE} (and \ac{SS}) methods often introduce distortions to the output signals, which were shown to degrade the performance of downstream tasks, e.g., \ac{ASR} \cite{ochiai2020beam}. To mitigate this issue, several works have demonstrated the advantage of combining \ac{TSE} (or \ac{SS}) with a back-end \ac{BF} \cite{heymann2016neural, erdogan2016improved, vzmolikova2019speakerbeam, ochiai2020beam, chen22e_interspeech}. The core idea is to use the output signals (or \ac{TF} masks) in the computation of the \acp{SCM} of a frequency-domain back-end \ac{BF}, e.g., \ac{MVDR}. A back-end \ac{MVDR} \ac{BF} enforces a distortionless constraint on the target speaker, which generally facilitates downstream tasks. In this work, we investigate extending the proposed \ac{BGTSE} method with a back-end \ac{MVDR} \ac{BF} following \cite{ochiai2020beam}. Specifically, we apply the proposed method to each channel independently, i.e., by changing the reference microphone $c$ in (\ref{eq:proposed}), and use the estimated time-domain waveforms directly to compute the \acp{SCM} of a back-end \ac{MVDR} \ac{BF}. Note that an estimate of the interferer signal is obtained by subtraction, i.e., $\widehat{x}^{(c)}_{N} = y^{(c)} - \widehat{x}^{(c)}_{S}$.

\vspace{-0.7em}
\section{Experimental Setup}
\label{sec:setup}

\vspace{-0.7em}
\subsection{Dataset}
\vspace{-0.1em}

We used simulated multi-channel $2$-speaker mixtures utilizing an extended version of the WHAMR! dataset  \cite{maciejewski2020whamr}. In particular, we extended the WHAMR! generation scripts to support an arbitrary microphone array geometry with its origin placed at the center of the dual microphone array defined in WHAMR!. In this work, we used a circular microphone array of $C=4$ elements having a radius sampled from $7.5$~cm to $12.5$~cm. For training the different models, we considered the subset of reverberant clean mixtures and used the reverberant sources as training targets, i.e., without dereverberation. The dataset consists of training, validation, and test splits of $20$k, $5$k, and $3$k examples, respectively. The reverberation time ranges from $0.1$~s to $1.0$~s and the source-to-array distance was randomly selected from $0.66$~m to $2.00$~m. The mixtures were created using a \ac{SIR} randomly sampled from $0$~dB to $5$~dB. The \textit{min} version of the dataset with a sampling frequency of $8$~kHz was used in all experiments. 


\vspace{-0.9em}
\subsection{Model Configurations}
\label{sec:model_configs}
\vspace{-0.1em}
The proposed method, shown in Figure \ref{fig:proposed}, consists of two blocks, a front-end \ac{BF} and a \ac{SC} \ac{TSE} block. In this work, we considered \ac{DSB}, \ac{SDB}, and \ac{MPDR} beamformer for the front-end \ac{BF}. The steering vector was computed assuming free-field and far-field. The front-end \ac{BF} was implemented in the \ac{STFT} domain with a window size of~$1024$~samples and a $75\%$ overlap. However, we note that, in the case of \ac{DSB}, it is possible to implement it in the time domain without the need for the \ac{STFT}. For the \ac{SC} \ac{TSE} block, we used the architecture provided in \cite{Delcroix2020}\footnote{We used the official implementation available Online \url{https://github.com/BUTSpeechFIT/speakerbeam}}, which utilizes a time-domain encoder-decoder structure based on the \ac{TCN} architecture \cite{Luo2019}.

Two \ac{SC} baselines were used in this study: an enrolment-based \ac{TSE} method, namely the \ac{TD-SpeakerBeam}\cite{Delcroix2020} and a \ac{SS} method using the \ac{TasNet} \cite{Luo2019}.
We also benchmarked with two \ac{MC} methods: MC-TasNet \cite{gu2019end}, which extends \ac{TasNet} to \ac{MC} input by using a parallel encoder, and the \ac{TSNF} \cite{gu2020temporal}, which utilizes spatial and directional features (6$\times$cosIPD + target speaker's \ac{AF}). For computing the spatial/directional features of the \ac{TSNF}, \ac{STFT} using 64-point FFT was used with a kernel size and stride equal to those of the time-domain encoder, following \cite{gu2020temporal}.

For a fair comparison, all methods employ the same \ac{TCN} architecture \cite{Luo2019}. For the encoder and decoder of all systems, we used a kernel size of~$16$~samples and a stride of~$8$~samples.  The hyperparameters of the \ac{TCN} architecture were chosen as follows: \mbox{$N=512$}, \mbox{$B=128$}, \mbox{$S_c=128$}, \mbox{$H=512$}, \mbox{$P=3$}, \mbox{$X=8$}, \mbox{$R=3$}, $\text{normalization}=\text{gLN}$, following the notation in \cite{Luo2019}. Similar to the TD-SpeakerBeam implementation, the auxiliary network in the proposed method consists of a time-domain encoder and a \ac{TCN} block with only one repeat (i.e., $R=1$), and the output embedding dimensionality $N_r=128$.

Furthermore, similar to \cite{ochiai2020beam}, we investigated combining the proposed method as well as the other baselines with a frequency domain back-end \ac{MVDR} \ac{BF}, as described in Section~\ref{sec:be_beamforming}. The back-end \ac{BF} was implemented in the \ac{STFT} domain using a window size of~$4096$~samples with an overlap of~$75\%$~to account for the large reverberation time. We also report the performance of the back-end \ac{BF} using the \ac{IRM} and the oracle signals.

\vspace{-0.7em}
\subsection{Training Setup}
For training the proposed and baseline methods, we used the \ac{SDR} loss function defined as, 
\begin{equation}
  \text{SDR}_\text{dB} := 10 \log_{10} \frac{\| x^{(c)}_S \|^2}{\| x^{(c)}_S -\widehat{x}^{(c)}_S \|^2}.
\end{equation}

\noindent
This is to ensure proper scaling of the estimated target signal, which is required in computing the \acp{SCM} of the back-end \ac{BF}. Adam optimizer \cite{Kingma2015} was used with an initial learning rate of~$10^{-3}$ and a weight decay of~$10^{-5}$. The maximum number of epochs was set to~$200$~and a batch size of~$6$~was used. The learning rate was halved if the validation loss did not decrease in~$3$~consecutive epochs. An early stopping patience of~$40$~epochs was used. The gradients were clipped if their $\ell_2$~norm exceeded a value~of~$5$. During training, the mixture signals (and enrolment for TD-SpeakerBeam) were cropped to~$4$~s. No dynamic mixing was applied. \Acl{PIT} \cite{kolbaek2017multitalker} was used in training the \ac{SS} baselines.

\vspace{-0.7em}
\section{Performance Evaluation}
\label{sec:results}
\vspace{-0.4em}
In this section, we first compare the performance of the proposed method with the baselines as well as their extension with a back-end \ac{BF}. Then, the generalization ability of the systems to unmatched input conditions is studied. Finally, we evaluate the robustness of the proposed method w.r.t. inaccurate input of the target speaker's \ac{DOA}. As an evaluation metric, we use the \ac{SI-SDR} \cite{leroux2019}, and report the improvement w.r.t. the input mixture. In the evaluation of all systems (without the back-end \ac{BF}), the first channel was used as the reference. Note that for evaluating the SS baselines, oracle selection was used to identify the target speaker \cite{Delcroix2020}.

{\bf Comparison with Baselines:}
Table~\ref{tab: main_results} shows the performance of the different systems (i.e., $\widehat{x}_{S}$) and their extension using a back-end \ac{BF} (i.e., $\widehat{x}_{S}^{\text{BF}}$). The \ac{MC} systems generally perform better than the \ac{SC} counterparts by taking advantage of the spatial properties of the sources. We can also observe that the proposed method using a \ac{DSB} as a front-end and \ac{TV} embeddings outperforms all \ac{SC} and \ac{MC} baselines. This shows the advantage of employing the front-end \ac{BF} to provide a correlated auxiliary signal with the target speaker as well as using \ac{TV} embeddings in a \ac{TSE} framework. Interestingly, the choice of the front-end \ac{BF} does not substantially alter the results, even though the \ac{SDB} and \ac{MPDR} are more spatially selective than \ac{DSB}. Although the front-end \ac{BF} provides limited enhancement of the target speaker ($\text{\ac{DSB}}=0.80$~dB, $\text{\ac{SDB}}=1.71$~dB, $\text{\ac{MPDR}}=1.93$~dB), this initial estimate is sufficient for identifying the target speaker. As expected, opting for \ac{TI} embeddings instead of \ac{TV} in the proposed method results in a drastic drop in performance by about $4.6$~dB. Attaching a back-end \ac{BF} provides an additional gain for all systems, where the proposed method still outperforms all baselines by at least $0.8$~dB. Remarkably, we can observe comparable performance between the oracle-mask \ac{MVDR} and the proposed method without the backend \ac{BF}, whereas with the back-end \ac{BF}, the proposed method achieves better scores and closes the gap with an oracle-signal \ac{MVDR}.

\begin{table}[t!]
\centering
\setlength{\tabcolsep}{1em} 
\renewcommand{\arraystretch}{1.05}
\caption{\ac{SI-SDR} [dB] of the systems (\text{$\widehat{x}_{S}$}) and their extension using a back-end \ac{BF} (\text{$\widehat{x}_{S}^{\text{BF}}$}).}
\vspace{-0.1em}
\label{tab: main_results}
\scalebox{0.9}{
\begin{tabular}{@{}lS[table-format=2.2]S[table-format=2.2]@{}}
\toprule
System                                        & \text{$\widehat{x}_{S}$}  	& \text{$\widehat{x}_{S}^{\text{BF}}$}  \\ \midrule
TD-SpeakerBeam \cite{Delcroix2020}            & 10.19          & 13.47          \\
TasNet \cite{Luo2019, ochiai2020beam}                         & 9.62          & 13.13	          \\ \midrule
MC-TasNet \cite{gu2019end, ochiai2020beam}               & 12.55          & 15.08          \\ 
\acs{TSNF} \cite{gu2020temporal}          & 12.26          &  15.20          \\ \midrule
Proposed ($\text{BF}_{\text{FE}}=\text{DSB}$, TV)           & 13.45          & \bf{16.02}          \\ 
Proposed ($\text{BF}_{\text{FE}}=\text{SDB}$, TV)       & 13.42          & 15.96          \\
Proposed ($\text{BF}_{\text{FE}}=\text{MPDR}$, TV)      & \bf{13.59}          & \bf{16.02}          \\ 
Proposed ($\text{BF}_{\text{FE}}=\text{DSB}$, TI)        &  8.80         & 12.25         \\ \midrule
Oracle \acs{IRM}                            & 11.38          & 13.60          \\ 
Oracle signal                          & $\infty$       & 21.29       \\ \bottomrule
\end{tabular}
}
\end{table}

\begin{table}[t!]
\centering
\setlength{\tabcolsep}{0.5em} 
\renewcommand{\arraystretch}{1.05}

\sisetup{detect-weight,mode=text}
\renewrobustcmd{\bfseries}{\fontseries{b}\selectfont}
\renewrobustcmd{\boldmath}{}
\newrobustcmd{\B}{\bfseries}

\caption{\ac{SI-SDR} [dB] of the systems without the back-end \ac{BF} on the different 2-speaker mixture conditions.}
\vspace{-0.1em}
\label{tab:generalization_input}
\scalebox{0.9}{
\begin{tabular}{@{}lS[table-format=2.2]S[table-format=2.2]S[table-format=2.2]S[table-format=2.2]@{}}
\toprule
         & \multicolumn{4}{c}{Condition} \\ \cmidrule(l){2-5}
System                                        & \text{A}  	& \text{\text{AN}} & \cellcolor{gray!20} \text{R}  	& \text{\text{RN}}  \\ \midrule

TD-SpeakerBeam \cite{Delcroix2020}                   & 13.38          & 6.01        & \cellcolor{gray!20} 10.19        & 5.61          \\
TasNet \cite{Luo2019}                                & 12.32          & 3.99        & \cellcolor{gray!20} 9.62         & 3.99	          \\ \midrule
MC-TasNet \cite{gu2019end}                           & 18.29          & 6.07        & \cellcolor{gray!20} 12.55        & 5.74         \\ 
\acs{TSNF} \cite{gu2020temporal}                     & 16.54          & 7.94        & \cellcolor{gray!20} 12.26        & 6.84         \\
Proposed ($\text{BF}_{\text{FE}}=\text{DSB}$, TV)    & \B 19.26     & \B 11.37  & \cellcolor{gray!20} \B 13.45   & \B 8.95 \\ \bottomrule
\end{tabular}
}
\vspace{-0.7em}
\end{table}

{\bf Generalization to Different Input Conditions:}
We assess the generalization ability by evaluating the trained models on the following unmatched $2$-speaker conditions: anechoic (A), anechoic+noise (AN), and reverberant+noise (RN) and compare with the matched reverberant (R) condition. These conditions correspond to different tasks in WHAMR! \cite{maciejewski2020whamr}. The performance of the systems without the back-end \ac{BF} is summarized in Table~\ref{tab:generalization_input}. It is evident from the results that the proposed method shows better generalization capability compared to the different baselines.

\begin{figure}[t!]
	\centering
	\includegraphics[width=0.95\columnwidth]{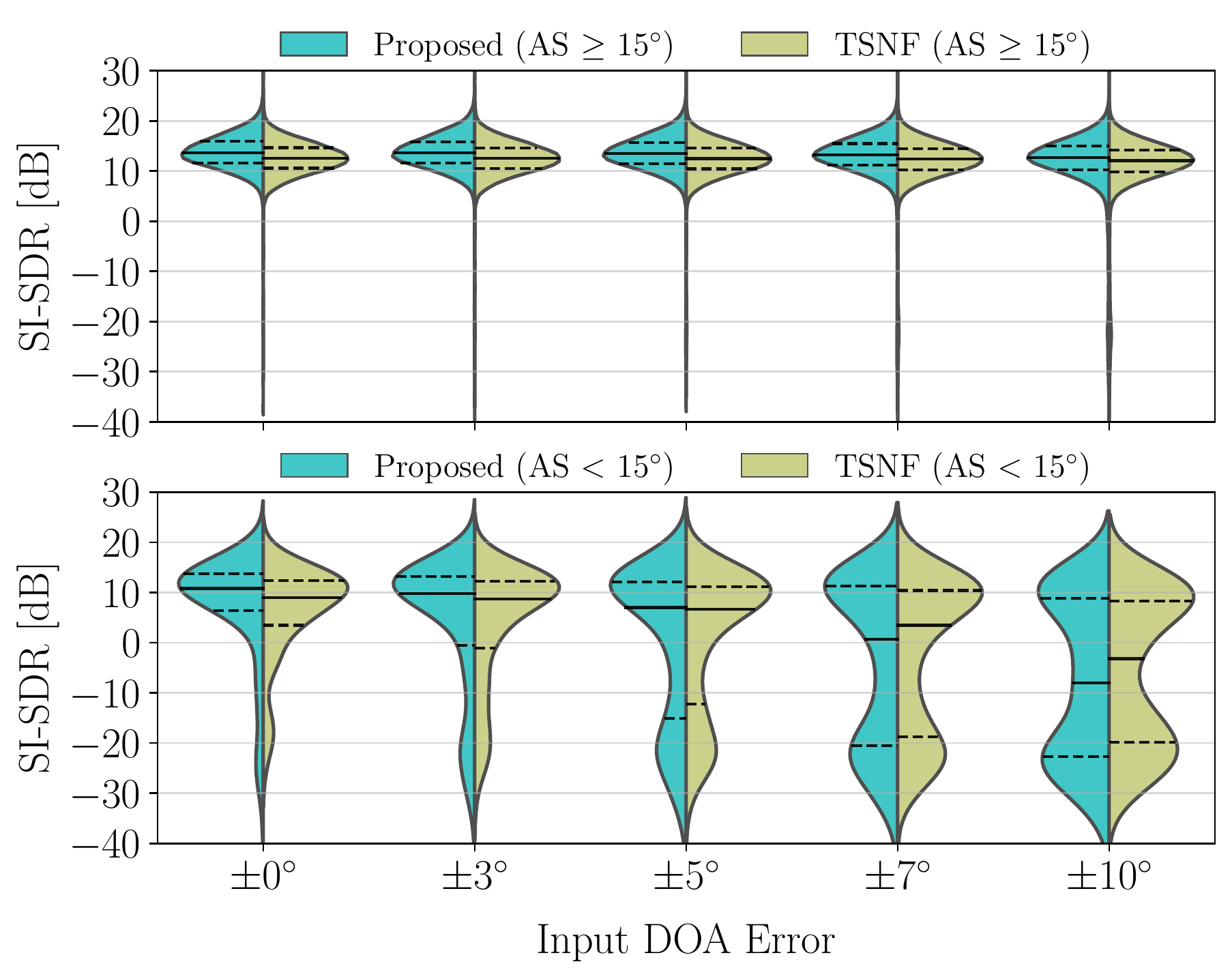}
	\vspace{-1em}
	\caption{Performance for inaccurate \ac{DOA}. (Top) \Acf{AS} between speakers larger than or equal to~$15^\circ$. (Bottom) \ac{AS} smaller than~$15^\circ$.
	}
	\label{fig:DOA_error}
 	\vspace{-1em}
\end{figure}

{\bf Effect of Erroneous DOA:}
In the proposed method and the \ac{TSNF} baseline, it is assumed that the \ac{DOA} of the target speaker is known. Here, we analyze the robustness of both systems against errors in the \ac{DOA} input for the matched reverberant condition (R). Figure~\ref{fig:DOA_error} shows the results of this analysis for both the proposed method ($\text{BF}_{\text{FE}}=\text{DSB}$, TV) and the \ac{TSNF} baseline. Note the results are reported without the back-end \ac{BF}.  For an \ac{AS}~$\ge 15^\circ$ between the two speakers, it can be seen that both methods are robust to errors in \ac{DOA} and that the proposed method exhibits slightly better performance than \ac{TSNF}. In contrast, for an \ac{AS}~$<15^\circ$, erroneous \ac{DOA} leads to worse scores for both systems. This behavior is expected since the target in both systems is only specified by the \ac{DOA}, and when the \ac{DOA} error becomes closer to the value of \ac{AS}, the systems tend to extract the speaker closer to the erroneous \ac{DOA}.   

\vspace{-0.7em}
\section{Conclusion}
\vspace{-0.5em}
In this paper, we introduced a \acs{BGTSE} method that consists of a \ac{SC} \acs{TSE} system guided by a front-end \ac{BF} steered towards the target speaker. The initial enhancement provided by the front-end \ac{BF} is capable of identifying the target speaker in the mixture. By allowing for \ac{TV} embeddings in the \ac{TSE} block to exploit the correspondence between the front-end \ac{BF} output and the target speech, the proposed \acs{BGTSE} method provides a significant improvement over several \ac{SC} and \ac{MC} baselines. In future work, we will explore different techniques in applying the proposed method in a causal/block-online fashion.


\vfill\pagebreak
\balance

\bibliographystyle{IEEEbib}
\bibliography{sapref}

\end{document}